# Surface *vs* bulk electronic structures of a moderately correlated topological insulator YbB$_6$ revealed by ARPES

N. Xu,[1,*] C. E. Matt,[1,2] E. Pomjakushina,[3] J. H. Dil,[4,1] G. Landolt,[1,5] J.-Z. Ma,[1,6] X. Shi,[1,6] R. S. Dhaka,[1,4,7] N. C. Plumb,[1] M. Radović,[1,8] V. N. Strocov,[1] T. K. Kim,[9] M. Hoesch,[9] K. Conder,[3] J. Mesot,[1,2,4] H. Ding,[6,10] and M. Shi[1,†]

[1]*Swiss Light Source, Paul Scherrer Institut, CH-5232 Villigen PSI, Switzerland*

[2]*Laboratory for Solid State Physics, ETH Zürich, CH-8093 Zürich, Switzerland*

[3]*Laboratory for Developments and Methods, Paul Scherrer Institut, CH-5232 Villigen PSI, Switzerland*

[4]*Institute of Condensed Matter Physics, Ecole Polytechnique Fédérale de Lausanne, CH-1015 Lausanne, Switzerland*

[5]*Physik-Institut, Universität Zürich, Winterthurerstrauss 190, CH-8057 Zürich, Switzerland*

[6]*Beijing National Laboratory for Condensed Matter Physics and Institute of Physics, Chinese Academy of Sciences, Beijing 100190, China*

[7]*Department of Physics, Indian Institute of Technology Delhi, Hauz Khas, New Delhi-110016, India*

[8]*SwissFEL, Paul Scherrer Institut, CH-5232 Villigen PSI, Switzerland*

[9]*Diamond Light Source, Harwell Science and Innovation Campus, Didcot OX11 0DE, UK,*

[10]*Collaborative Innovation Center of Quantum Matter, Beijing, China*

**Systematic angle-resolved photoemission spectroscopy (ARPES) experiments have been carried out to investigate the bulk and (100) surface electronic structures of a topological mixed-valence insulator candidate, YbB$_6$. The bulk states of YbB$_6$ were probed with bulk-sensitive soft X-ray ARPES, which show strong three-dimensionality as required by cubic symmetry. Surprisingly the measured Yb 4*f* states are located at ~ 1 eV and 2.3 eV below the Fermi level ($E_F$), instead of being near $E_F$ as indicated by first principle calculations. The dispersive bands near $E_F$ are B *2p* states, which hybridize with the 4*f* states. Using surface-sensitive vacuum-ultraviolet ARPES, we revealed two-dimensional surface states which form three electron-like Fermi surfaces (FSs) with Dirac-cone-like dispersions. The odd number of surface FSs gives the first indication that YbB$_6$ is a moderately correlated topological insulator. The spin-resolved ARPES measurements provide further evidence that these surface states are spin polarized with spin locked to the crystal momentum. We have observed a second set of surface states with different topology (hole-like pockets). Clear folding of the bands suggest their origin from a 1×2 reconstructed surface. The topological property of the reconstructed surface states needs further studies.**

Recently the study of topological nontrivial surface (edge) states with strong electron interactions [1-8] has become one of the focuses in the research of topological insulators (TIs) [9-10], particular efforts have been dedicated to realize new TIs which contain rare-earth elements. Intensive experimental investigations on $SmB_6$ have provided compelling evidences that topologically non-trivial surface states can be hosted in the Kondo insulating phase of the bulk at low temperatures [11-26], consistent with the initial theoretical prediction that $SmB_6$ is a topological Kondo insulator [1-5]. Theoretical considerations [27] have also indicated that, due to the inversion of 5*d* with 4*f* bands at the *X* points in the bulk Brillouin zone (BZ) (Fig. 1b), topological surface states protected by time reversal symmetry (TRS) should appear at the surface of the bulk insulating phase of $YbB_6$ which is iso-structural to $SmB_6$ (Fig. 1a). However, in difference to the Kondo insulator $SmB_6$ for which the bulk insulating phase stems from Kondo effect at low temperatures, it has been shown by band structure calculations [27] that the occupation number of the *f*-orbitals is close to 14 in $YbB_6$. Therefore, $YbB_6$ is close to the mixed valence boundary and the possible topological surface states appear in an intermediately correlated band insulator in bulk. It will be of great importance to investigate the electronic structure and further reveal the topological properties of $YbB_6$.

Using the combination of soft x-ray ARPES (SX-ARPES) and conventional ARPES with vacuum ultraviolet light as the excitation source (VUV-ARPES) as well as spin-resolved ARPES (SARPES), we pursued a systematic study of the bulk and surface electronic states of $YbB_6$. The delocalized bulk states near $E_F$ are B-*2p* states, exhibiting highly three-dimensionality and consistent with cubic crystal symmetry of the system. However, the observed localized Yb-*4f* states are very different to the one predicted by the first-principle calculations [27], located at ~ 1 and 2.3 eV below the $E_F$ instead of being near $E_F$. The two-dimensional surface states (electron-like) are observed on the *n-type* surface, forming three Fermi surfaces, one is around the surface Brillouin zone (SBZ) center $\bar{\Gamma}$ point and two are centered at the mid-point of the SBZ boundary ($\bar{X}$ points). These surface states exhibit Dirac-cone-like dispersions with the Dirac point (DP) estimated at 0.49 and 0.36 eV below $E_F$. The SARPES results indicate that the surface states at the $\bar{X}$ points are spin polarized in plane and locked to the crystal momentum, consistent with the topological insulating states in YbB$_6$. Our ARPES results give compelling spectroscopy evidences that YbB$_6$ is an intermediately correlated topological insulator. It should be mentioned that we have also observed another set of surface states (hole-like) on the *p-type* surface. Clear folding bands at the $\bar{M}$ points are observed, suggests a 1×2 reconstruction happened at this *p-type* (100) surface.

Figure 1c shows the angle-integrated photoemission spectra of YbB$_6$ taken with the VUV light (35 eV) and soft X-ray (1000 eV), as well as the spectrum of SmB$_6$ for comparison. Due to the increased bulk-sensitivity, the SX-ARPES mainly probes the bulk states, which is manifested by the well-defined symmetrical spectral peaks of the *4f* states. We measured different samples as well as different positions on one cleaving surface, and found they are quite homogenous without phase separation. In contrast, the ARPES spectra taken with $h\nu$ = 35 eV shows two sets of features strongly depend on surface positions. For one type of surface (named as *n-type* surface), the spectra shows multi-peak structure with significant spectral weight shifted to higher binding energies by about 0.15 eV relative to the bulk ones. This shift is attributed to the surface contribution because the short escape-depth of photoelectrons in the VUV regime, which is consistent with that the VUV-ARPES is more sensitive to the surface states than SX-ARPES [28]. For the other type of surfaces states, the *4f* states peaks shift towards low binding energies about 0.25 eV, therefore we name this type of surface as *p-type* surface. We notice that the binding energies of Yb *4f* states both in the bulk and at the surface are comparable to that of the *f* electrons in pure Yb ($E_B$ = 1.3 and 2.5 eV). On the other hand, the measured *f* states are very different to the calculated one [26], as well as that in its iso-structure compound SmB$_6$, in which one *4f* state resides a few tens meV below $E_F$.

**Bulk electronic structure of YbB$_6$.** In determining the bulk electronic structure of YbB$_6$, we carried out SX-ARPES measurements with large escape-depth of the photoelectrons in the photon energy range of 600 - 1100 eV to cover a few bulk BZs in momentum space. Figures 2a and 2b show the band dispersions along the high symmetry lines *X-Γ-X* and *X-M-X*, taken at $k_z$ ~ 18π and 19π ($h\nu$ = 700 and 800 eV), respectively. The corresponding energy distribution curves (EDC) plots are presented in Figs. 2c and 2d.

The nearly non-dispersive intense spectral peaks are the bulk Yb-*4f* states, observed at $E_B$ ~ 1 eV and 2.3 eV. Two sharp and highly dispersive bands, the γ and δ bands, are clearly visible in the energy range from $E_B$ ~ -3.5 eV to $E_F$, which hybridize with the *4f* states. The top and bottom of the γ band locates at the *X* and *M* point, respectively. The intensity decreases when the band approaches to $E_F$, and the band gap (energy distance from the band top to $E_F$) is estimated to be 80 meV, consistent with the band gap size observed in optical data [29]. By comparing the dispersive feature and the band top/bottom location in *k*-space with the calculations [27], we attribute the γ band to the B-*2p* band. The other dispersive band δ is far below $E_F$, the band top and bottom locate at the Γ and *X* points at binding energies ~ 1 eV and ~ 3.5 eV, respectively. To confirm that the measured γ and δ bands represent the bulk electronic structure of YbB$_6$, in Figs. 2e-2g and Figs. 2h-2j we plot the ARPES intensity maps in the symmetrically equivalent $k_x$-$k_y$ and $k_y$-$k_z$ planes at the constant energies E1, E2 and E3 (0.15, 1.55 and 2.65 eV below $E_F$) as indicated in Figs. 2a-2b. One can immediately realize that the γ and δ bands are highly three-dimensional and the intensity pattern is consistent with the cubic symmetry of the bulk crystal structure, thus affirming their bulk origin. The unexpected bulk electronic structure where the 4*f* states hybridize with the 2*p* states at higher binding energies, suggests the *f*-orbitals is fully occupied, and indicates that YbB$_6$ is out of the mixed-valence region resulting in Yb valence of +2.

**The surface states topology on the *n-type* (100) surface of YbB$_6$.** To explore the possible topological surface states in YbB$_6$ we used more surface-sensitive VUV-ARPES (as indicated in Fig. 1c) with the photon energy range of 30-90 eV to investigate the electronic structure on the cleaved (100) surface. Different to SX-ARPES results, we observed a few energy bands crossing $E_F$ on the *n-type* surfaces. As shown in Fig. 3b, these bands form three FSs, one circle is around the SBZ center ($\bar{\Gamma}$) and two ellipses are centered at the high symmetry points $\bar{X}$ at the SBZ boundary with the long axes along the $\bar{\Gamma}$-$\bar{X}$ direction. To trace the detailed dispersive features of these energy bands, in Figs. 3e and 3g we plot the ARPES spectra along the high symmetry lines $\bar{\Gamma}$-$\bar{X}$ and $\bar{X}$-$\bar{M}$ directions, with corresponding energy distribution curves (EDC) plots shown in Figs. 3f and 3h. Near the SBZ center $\bar{\Gamma}$, the energy band ($\alpha$ band) shows a Dirac-cone-like linear dispersive feature (Figs. 3e-3f) with the extrapolated DP (around 0.49 eV below $E_F$) deep into the valence band (VB). Similarly, the energy band at the $\bar{X}$ points ($\beta$ band) also shows a Dirac-cone-like dispersive feature, with the DP located at 0.36 eV below $E_F$, which can also seen in Figs. 3e-3h. The $k_F$ and Fermi velocity ($v_F$) for the $\alpha$ band are 0.14 (Å$^{-1}$) and 3.5 (eV*Å), and for the $\beta$ band they are 0.12/0.17 (Å$^{-1}$) and 3.0/2.1 (eV*Å) for the $\bar{\Gamma}$-$\bar{X}$ and $\bar{X}$-$\bar{M}$ directions, respectively. We notice that a conduction band (CB) is observed near $E_F$ around the $\bar{X}$ points (Figs. 3e-3j).

To further confirm that the metallic states (the $\alpha$ and $\beta$ bands) are surface bands, we have carried out an ARPES measurement along the $\bar{\Gamma}$-$\bar{X}$ direction for different $k_z^B$ values by tuning the photon energy. In Figs. 3a we plot ARPES spectra with $k_{//}$ oriented along the $\bar{\Gamma}$-$\bar{X}$ direction taken with different photon energies from 30 to 70 eV. One can see that, although the spectral weights of the $\alpha$ and $\beta$ bands vary with photon energy due to photoemission matrix element effects, the dispersions of the $\alpha$ and $\beta$ bands stay fixed. As also seen from the $h\nu$-$k_{//}$ FS intensity plot in Fig. 3d, one immediately recognizes that the $\alpha$ and $\beta$ bands form two-dimensional FSs in the $h\nu$-$k_{//}$ plane, in contrast to the bulk bands which show strongly three-dimensionality as seen in Figs. 2h-2j. In fact, when we plot the extracted dispersions for different photon energies in Fig. 3c, their linear dispersions overlap each other within the experimental uncertainties, demonstrating the two-dimensional nature of the $\alpha$ and $\beta$ bands. This two-dimensional feature is different from the bulk $\gamma$ and $\delta$ bands, indicating the surface origin of the $\alpha$ and $\beta$ bands.

**Spin polarization measurements of the *n-type* surface states.** From the dispersions of the $\alpha$ and $\beta$ bands (Figs. 3) we see that the FSs of these surface states in YbB$_6$ are formed by electron-pockets around $\bar{\Gamma}$ and $\bar{X}$ points (Fig. 3a). The number, the shapes and the locations of these surface FSs show great similarity to that of SmB$_6$ [7-9], the first realized TKI, as well as to the theoretically predicted topological surface states of YbB$_6$ [26]. The odd number of surface FSs around the Kramer's points and the Dirac-cone-like dispersions suggest the possibility that the $\alpha$ and $\beta$ bands are topologically non-trivial surface states. In order to examine the topological nature of the surface states, we investigated the spin polarization of the low-energy states of the YbB$_6$ samples directly by SARPES. Considering the low efficiency of the SARPES measurements, we focus on the $\beta$ band which has a much stronger intensity as seen in Figs. 3e-3j. In Fig. 4b, we plot the spin-integrated ARPES intensity near the $\bar{X}$ point (as indicated by the green solid line in Fig. 4a), with the exactly same experimental condition as the SARPES measurements discussed as following. One can realize that although the low resolution give a much reduced detail compared to the high resolution data in Fig. 2, the surface states $\beta$ can be clearly observed due to the large band width. It also manifests that the intensity is highly dominated by the surface states $\beta$ band near the Fermi level because of the much deeper position of the *4f* states. Figures 4c shows the spin resolved MDC intensity $I^{\uparrow\downarrow}_x$ for the *x* direction (as indicated in the coordinate system in Fig. 4a), measured at $E_{SR}$ (0.1 eV below $E_F$, as indicated in Fig. 4b) with $h\nu = 35$ eV and linearly horizontal polarized light. As shown in Fig. 4c, there is a clear difference in $I^{\uparrow}_x$ and $I^{\downarrow}_x$. The peaks of the $I^{\uparrow}_x$ and $I^{\downarrow}_x$ MDCs corresponds to the two branches of the $\beta$ bands, indicating that the surface state bands are spin-polarized along the *x* direction. It is clearly seen in the net spin polarization MDC plot in Fig. 4d. In contrast we observed a near zero spin polarization in the out-of-plane direction.

The observed spin polarization of the surface states (Figs. 4), together with their Dirac-cone-like dispersions (Figs. 3), offer compelling spectroscopy evidences that YbB$_6$ is a moderately correlated topological insulator. However, since the Yb *4f* states in YbB$_6$ are far below the chemical potential as indicated (Fig. 1c and Figs. 2a-2d), it is unlikely that the topological surface states in YbB$_6$ are related to the *d-f* band inversion at the *X* point as it was anticipated in theoretical considerations [26]. Here we propose another possibility, -i.e. that the non-trivial surface states are associated with the bulk *p-d* band inversion at the *X* point. As suggested by the first-principles calculations, the band bottom of the Yb *5d* band is always below $E_F$ and overlaps with the B *2p* band. The hybridization between the *p* and *d* band opens a band gap around the chemical potential. Due to bulk *p-d* band inversion at the *X* point, the Dirac-cone-like dispersive topological surface states occur to fulfill the symmetry requirements of the system.

**The surface states topology on the *p-type* (100) surface of YbB$_6$.** Besides the *n-type* surface with electron-like topological surface states (Figs. 3-4), in some locations of the (100) surface of all the cleaved samples, we have always observed a second set of surface states. These FSs are formed by a hole-pocket (named as $α_h$ band) at the SBZ center the $\bar{Γ}$ point, as well as the other hole-like band (named as $β_h$ band) at the $\bar{X}$ point with band top close to $E_F$ (see Figs. 5e-5j). The photon energy dependence measurements (Figs. 5a and 5d) indicate the $α_h$ and $β_h$ bands exhibit two-dimensionality, originating from the surface states. We notice that there are some additional intensities around the $\bar{M}$ points in the FS mapping shown in Fig. 5a. The corresponding bands at the $\bar{M}$ point can be clearly resolved in Figs. 5f and 5j. We notice the dispersions of the bands at the $\bar{M}$ point show great similarity as the one at the $\bar{X}$ point. This is better illustrated in the band structure configuration shown in Figs. 5g and 5h. The inner band at the $\bar{M}$ point corresponds to the $β_p$ band along $\bar{X}$-$\bar{M}$ direction (Fig. 5f-5h) with a wave-vector $(0, π)$, and the outer band corresponds to the $β_h$ band along $\bar{Γ}$-$\bar{X}$ direction (Fig. 5e-5g) with a wave-vector $(π, 0)$. It is further confirmed by the ARPES intensity maps at the constant energies $E_B = 0.1$ eV in Fig. 5c. The two-crossing-ellipse shaped intensity pattern at the $\bar{M}$ point and the ellipse shaped pattern at the $\bar{X}$ point are well connected by $(π, 0)$ and $(0, π)$ wave-vectors. This suggests that the bands at the $\bar{M}$ point originate from the folded $β_h$ band from the $\bar{X}$ point, due to the superposition of (1×2) and (2x1) surface reconstructions on the *p-type* surface. To clarify this idea one would need to apply spatial-resolved spectroscopy, e.g. STM/STS or PEEM. We notice that the odd number of FSs of the reconstructed *p-type* surface states, especially the $α_p$ band could also be topologically non-trivial. This is supported by that the topologically non-trivial structures are encoded in the bulk band structure, and are independent of the detailed surface conditions. However, the weak intensity for the $α_p$ band makes it extremely challenging to conduct SARPES measurements.

**Conclusion.** Our systematic SX-ARPES study, gives a comprehensive description of the bulk electronic structures of YbB$_6$, constitute a foundation to examine and understand the electronic and possibly topological properties in YbB$_6$. The large discrepancy between the experimentally determined and the calculated band structures is unexpected and indicates the *f*-orbitals is fully occupied, and Yb is in bivalence states in YbB$_6$, instead of mixed-valence Sm in SmB$_6$. The VUV-ARPES results indicate that the electron-like metallic surface states are topological surface states, with spin-polarized in plane and locked to the crystal momentum as revealed by SARPES measurements. The additional hole-like metallic surface states are always observed, and the clear folding bands indicate a (1×2) reconstruction happened on this type of surface. Our SX-ARPES, VUV-ARPES as well as SARPES results give a comprehensive description on the bulk and surface electronic structures in YbB$_6$, and offer compelling spectroscopy evidences that YbB$_6$ is a moderately correlated topological insulator.

**METHODS**

VUV ARPES measurements were performed at Surface and Interface (SIS) beamline at Swiss Light Source (SLS), and the beamline I05-ARPES at Diamond with the energy and angular resolutions 5-15 meV and 0.2°, respectively. SARPES measurements were also performed at SIS beamline at the SLS with the COPHEE [30] station using two 40 kV classical Mott detectors, and with photon energies 35 eV and linear light polarizations. The temperature for VUV-ARPES and SARPES measurements shown in the paper was set to 20 K. Soft X-ray ARPES measurements were performed at the Advanced Resonant Spectroscopies (ADRESS) beamline at SLS, and data were collected using circular-polarized light with an overall energy resolution on the order of 50-80 meV, at $T = 10$ K. High-quality single crystals of YbB$_6$ were grown by flux method. Samples were cleaved *in-situ* along the (001) crystal plane in an ultrahigh vacuum better than $2\times10^{-10}$ Torr. A shiny mirror-like surface was obtained after cleaving the samples, confirming their high quality. The Fermi level of the samples was referenced to that of a gold film evaporated onto the sample holder.

15. Neupane, M. *et al*. Surface electronic structure of the topological Kondo-insulator candidate correlated electron system $SmB_6$. *Nat. Commun.* 4:2991 (2013).

16. Jiang, J. *et al*. Observation of possible topological in-gap surface states in the Kondo insulator $SmB_6$ by photoemission. *Nat. Commun.* 4:3010 (2013).

17. Xu, N. *et al*. Direct observation of the spin texture in strongly correlated $SmB_6$ and experimental realization of the first topological Kondo insulator. Submitted. (2013)

18. Li, G. *et al*. Quantum oscillations in Kondo insulator $SmB_6$. arXiv.org (2013), 1306.5221v1.

19. Kim, D.J., Xia, J., & Fisk, Z., Topological surface state in the Kondo insulator samarium hexaboride. arXiv.org (2013), 1307.0448.

20. Chen, F. *et al*. Angular dependent magnetoresistance evidence for robust surface state in Kondo insulator $SmB_6$. arXiv.org (2013), 1309.2378.

21. Yue, Z.-J., Wang, X.-L., Wang, D.-L., Dou, S.-X., Wang J.-Y., Four-fold symmetric magnetoresistance in Kondo insulator $SmB_6$. arXiv.org (2013), 1309.3005.

22. Thomas *et al*. Weak antilocalization and linear magnetoresistance in the surface state of $SmB_6$. arXiv.org (2013), 1307.4133.

23. Yee, M.M. *et al*. Imaging the Kondo insulating gap on $SmB_6$. arXiv.org (2013), 1308.1085v1.

24. Ruan, W. *et al*. Emergence of a coherent in-gap state in SmB6 Kondo insulator revealed by scanning tunneling spectroscopy. arXiv.org (2014), 1403.0091.

25. Min, C.-H., *et al*., The importance of Charge Fluctuations for the Topological Phase in $SmB_6$. arXiv.org (2013), 1312.1834.

26. Denlinger, J. D. *et al*., Temperature Dependence of Linked Gap and Surface State Evolution in the Mixed Valent Topological Insulator $SmB_6$. arXiv.org (2013), 1312.6637.

27. Weng, H.M., Zhao, J.Z., Wang, Z.J., Fang, Z. and Dai, X., Topological Crystalline Kondo Insulator in Mixed Valence Ytterbium Borides. *Phys. Rev. Lett.* 112, 016403 (2014).

28. Razzoli, E. *et al*., Bulk Electronic Structure of Superconducting $LaRu_2P_2$ Single Crystals Measured by Soft-X-Ray Angle-Resolved Photoemission Spectroscopy. *Phys. Rev. Lett.* 108, 257005 (2012).

**Acknowledgements**

This work was supported by the Sino-Swiss Science and Technology Cooperation (Project No. IZLCZ2138954), the Swiss National Science Foundation (Grant No. 200021-137783), and MOST (Grant No. 2010CB923000) and NSFC. We thank staff of beamline I05-ARPES at Diamond and the SIS and ADRESS beamline at SLS for their excellent support.

**Correspondence** and requests for materials should be addressed to N.X. (e-mail: nan.xu@psi.ch) or M.S. (e-mail: ming.shi@psi.ch).

**Figure Captures:**

**Fig. 1 |** a, The CsCl-type structure of YbB$_6$ with $Pm\overline{3}m$ space group. b, The first Brillouin zone of YbB$_6$ and the projection on the cleaving surface. High-symmetry points are also indicated. c, ARPES measured Yb *4f* states spectrum for bulk states with incident photon energy $h\nu$ = 1000 eV, as well as for the surface states on the *n-type* and *p-type* (100) surface of YbB$_6$ with $h\nu$ = 35 eV. The spectrum of SmB$_6$ is also shown for comparison.

**Fig. 2 | Three-dimensional bulk electronic structure of YbB$_6$ revealed by soft X-ray ARPES.** a-b, ARPES intensity along high symmetry lines *X-Γ-X* and *M-X-M* taken at $k_x$ = 2π and $k_z$ around 18π ($h\nu$ = 700 eV) and 19π ($h\nu$ = 800 eV), respectively. c-d, Corresponding EDCs plots. e-g, In plane ($k_x$-$k_y$) ARPES intensity mapping at binding energy at E1, E2 and E3 (0.15, 1.55 and 2.65 eV below $E_F$) for YbB$_6$ measured at $h\nu$ = 700 eV, corresponding to $k_z$ ~ 18π. h-j, Out of plane ($k_y$-$k_z$) ARPES intensity mapping at binding energy at E1, E2 and E3 for YbB$_6$ measured with photon energy from 600 to 1080 eV, corresponding to $k_z$ from 16π to 22π. The lines in e-j are guides to the eyes to trace the pockets.

**Fig. 3 | Dirac-cone-like surface states on the *n-type* surface of YbB$_6$.** a, Near-$E_F$ ARPES intensity measured with photon energy from 30 to 70 eV as a function of wave vector and $E_B$ along the high symmetry cut $\overline{\Gamma}$-$\overline{X}$. b, In plane ($k_x$-$k_y$) ARPES intensity mapping at $E_F$ for YbB$_6$, measured at $h\nu$ = 35 eV; this intensity is obtained by integrating the spectrum within ±5 meV of $E_F$. c, Extracted dispersions of the α and β bands for different photon energies. d, Plot of the ARPES intensity at $E_F$ along the high symmetry cut $\overline{\Gamma}$-$\overline{X}$ as a function of photon energies from 30 to 90 eV. e, ARPES intensity plot along the $\overline{\Gamma}$-$\overline{X}$ direction with $h\nu$ = 35 eV. f, Corresponding EDC plot. g-h, same as e-f but for the $\overline{X}$-$\overline{M}$ direction. All the results shown in Fig. 2 are obtained with right-handed circular polarized light (C+).

**Fig. 4 | SARPES measurements on the surface states on the *n-type* surface of YbB$_6$.** a, Schematic showing the Fermi surface of the surface states on the *n-type* surface of YbB$_6$,

with the green line indicating the locations of spin measurements in the $k_x$-$k_y$ plane. The circle at the BZ center $\bar{\Gamma}$ point and ellipses at the middle points of BZ boundaries ($\bar{X}$ points) indicate the Fermi surfaces of the α and β bands, respectively. b, ARPES intensity near the $\bar{X}$ point taken with exactly same experimental condition as the SARPES measurements. c, Measured spin-resolved intensity (MDC) projected along the *x* direction for the surface states β band at $E_{SR}$. The red and blue symbols are the intensity of spin up and spin down states, respectively. The energy is chosen at $E_{SR}$ indicated by the green line in a, 0.1 eV below the $E_F$. c, Corresponding spin polarization for the surface states measured at the $E_{SR}$. All the results shown in Fig. 4 are obtained with linear horizontal polarized light, $h\nu = 35$ eV.

**Fig. 5 | Surface states topology on the *p-type* surface of YbB$_6$.** a, Near-$E_F$ ARPES intensity measured with photon energy from 30 to 70 eV as a function of wave vector and $E_B$ along the high symmetry cut $\bar{\Gamma}$-$\bar{X}$. b, In plane ($k_x$-$k_y$) ARPES intensity mapping at $E_F$ for YbB$_6$, measured at $h\nu = 35$ eV; this intensity is obtained by integrating the spectrum within ±5 meV of $E_F$. c, Similar as b but taken at $E_B = 0.1$ eV. d, Plot of the ARPES intensity at $E_F$ along the high symmetry cut $\bar{\Gamma}$-$\bar{X}$ as a function of photon energies from 30 to 90 eV. e, ARPES intensity plot along the $\bar{\Gamma}$-$\bar{X}$ direction with $h\nu = 35$ eV. f, Same as e but along the $\bar{X}$-$\bar{M}$ direction. g-h, The band illustration along the high symmetry cuts $\bar{\Gamma}$-$\bar{X}$ and $\bar{X}$-$\bar{M}$. i-j, Corresponding EDC plots along the high symmetry cuts $\bar{\Gamma}$-$\bar{X}$ and $\bar{X}$-$\bar{M}$. All the results shown in Fig. 5 are obtained with right-handed circular polarized light (C+).

Figure 1

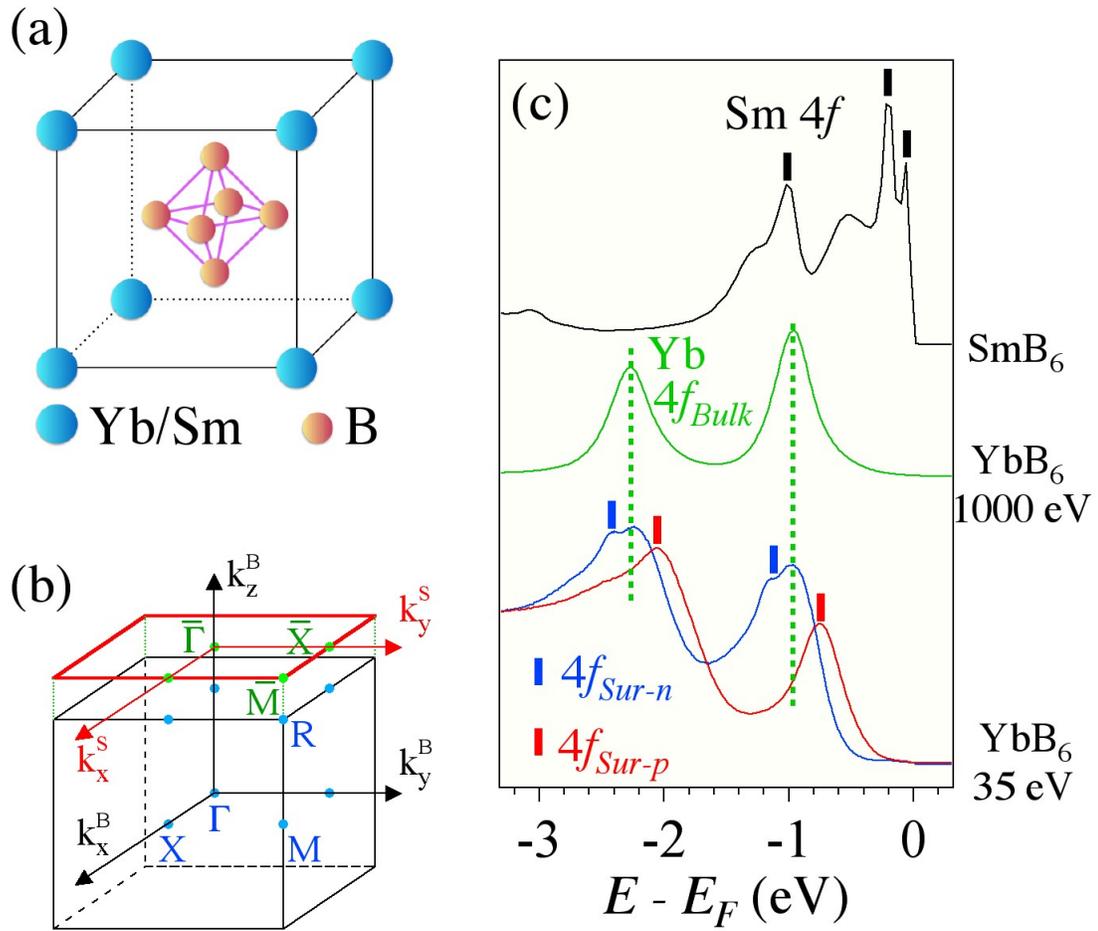

Figure 2

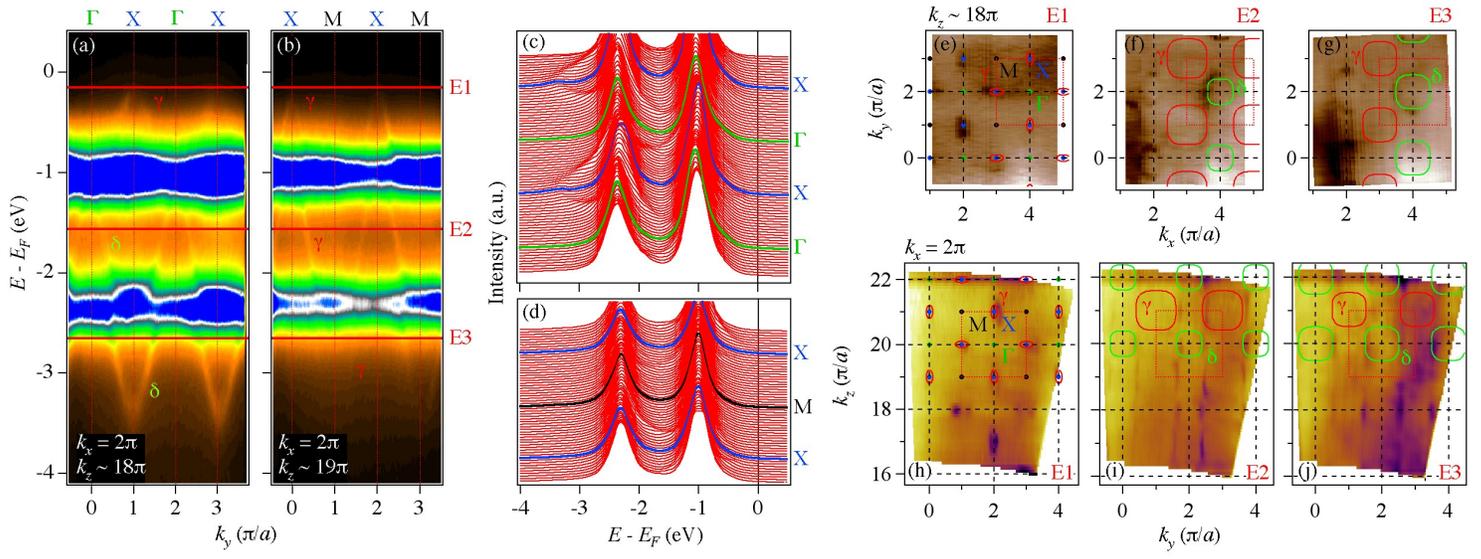

Figure 3

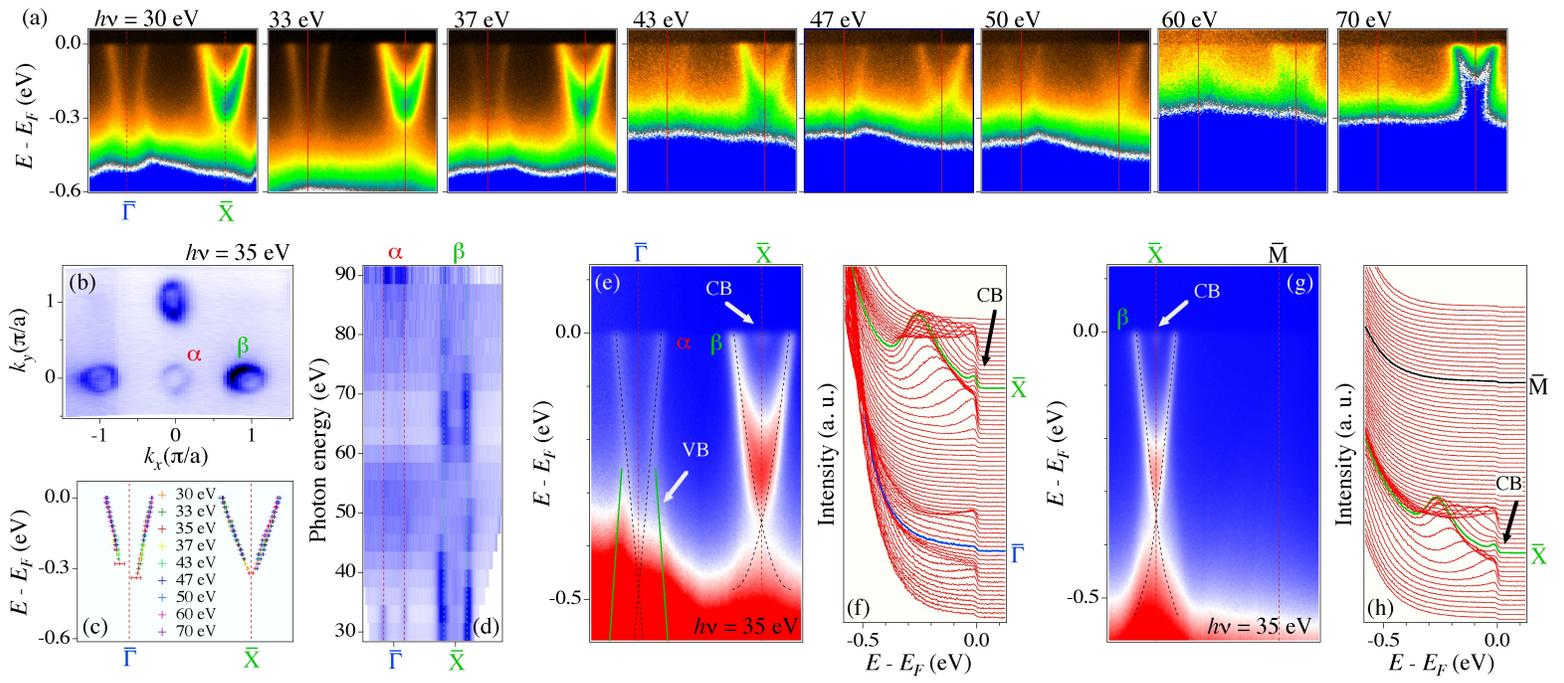

Figure 4

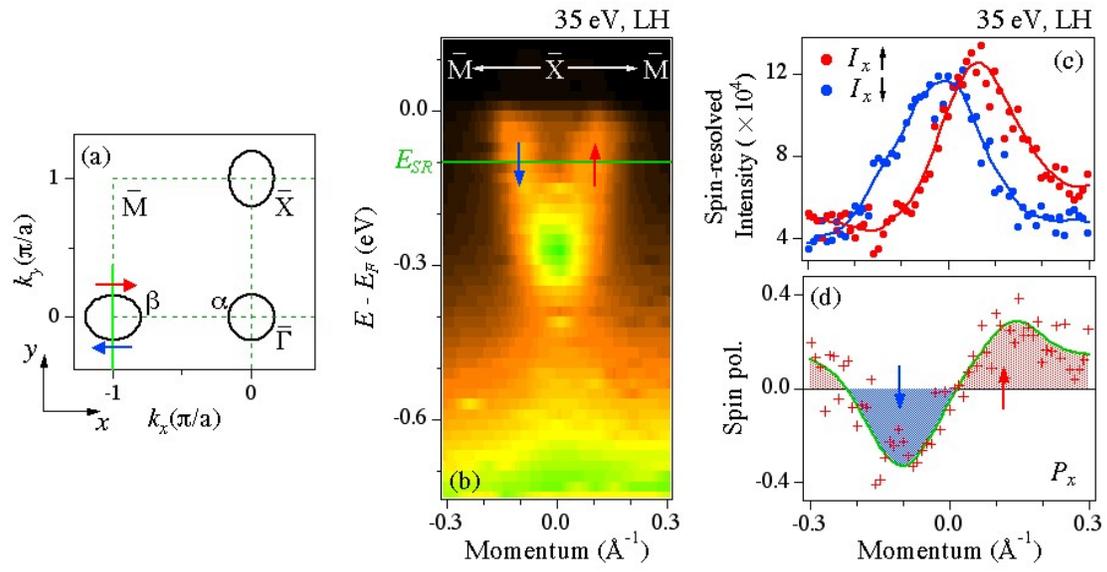

Figure 5

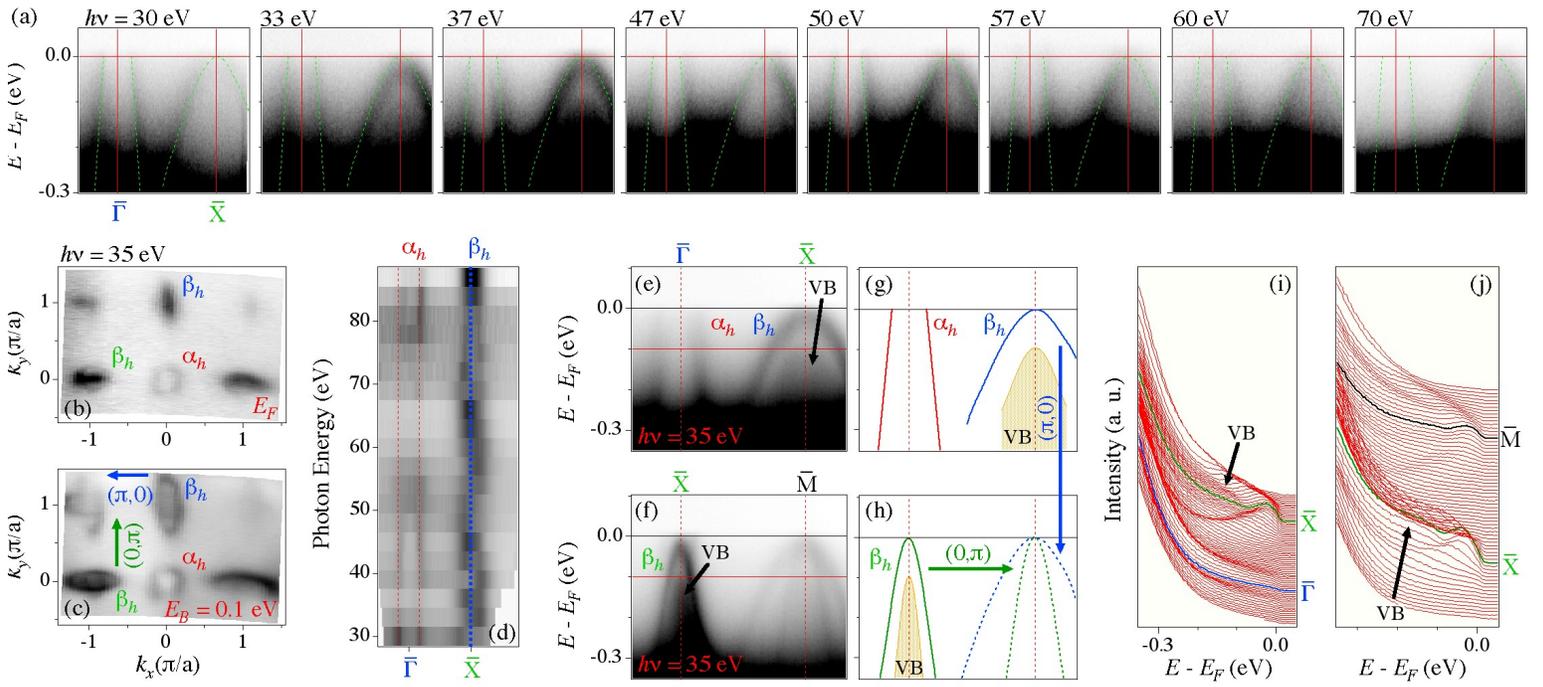